\documentclass[runningheads]{llncs}

\usepackage[T1]{fontenc}
\usepackage{graphicx}
\usepackage{tabularx}
\usepackage{soul}
\usepackage{lscape}
\usepackage{listings}
\usepackage{xcolor}
\usepackage{amsmath}
\usepackage{subcaption}
\usepackage{enumerate}
\usepackage{comment}

\title{AI Cards: Towards an Applied Framework for Machine-Readable AI and Risk Documentation Inspired by the EU AI Act}

\begin{document}

\title{AI Cards: Towards an Applied Framework for Machine-Readable AI and Risk Documentation Inspired by the EU AI Act}

\titlerunning{AI Cards}

\author{Delaram Golpayegani\inst{1}\orcidID{0000-0002-1208-186X} \and \newline
Isabelle Hupont\inst{2}\orcidID{ 0000-0002-9811-9397} \and
Cecilia Panigutti\inst{2}\orcidID{0000-0002-6552-787X} \and
Harshvardhan J. Pandit\inst{3}\orcidID{0000-0002-5068-3714}\and \newline
Sven Schade\inst{2}\orcidID{0000-0001-5677-5209} \and
Declan O'Sullivan\inst{1}\orcidID{0000-0003-1090-3548} \and
Dave Lewis\inst{1}\orcidID{0000-0002-3503-4644} }
\authorrunning{D. Golpayegani et al.}

\institute{ADAPT Centre, Trinity College Dublin, Dublin, Ireland 
\email{\{sgolpays,declan.osullivan,delewis\}@tcd.ie} \\ \and
European Commission, Joint Research Centre (JRC), Ispra, Italy \and
ADAPT Centre, Dublin City University, Dublin, Ireland \email{me@harshp.com}}
\maketitle              
\begin{abstract}
 With the upcoming enforcement of the EU AI Act, documentation of high-risk AI systems and their risk management information will become a legal requirement playing a pivotal role in demonstration of compliance. Despite its importance, there is a lack of standards and guidelines to assist with drawing up AI and risk documentation aligned with the AI Act. This paper aims to address this gap by providing an in-depth analysis of the AI Act's provisions regarding technical documentation, wherein we particularly focus on AI risk management. On the basis of this analysis, we propose \textbf{AI Cards} as a novel holistic framework for representing a given intended use of an AI system by encompassing information regarding technical specifications, context of use, and risk management, both in human- and machine-readable formats. While the human-readable representation of AI Cards provides AI stakeholders with a transparent and comprehensible overview of the AI use case, its machine-readable specification leverages on state of the art Semantic Web technologies to embody the interoperability needed for exchanging documentation within the AI value chain. This brings the flexibility required for reflecting changes applied to the AI system and its context, provides the scalability needed to accommodate potential amendments to legal requirements, and enables development of automated tools to assist with legal compliance and conformity assessment tasks. To solidify the benefits, we provide an exemplar AI Card for an AI-based student proctoring system and further discuss its potential applications within and beyond the context of the AI Act. 

\end{abstract}
\textcolor{red}{This work has been accepted and will be presented at the Annual Privacy Forum (APF) 2024 in Karlstad, Sweden, 4-5 September, 2024. This contribution will be published in the APF 2024 Springer LNCS proceedings.}

\section{Introduction}

Artificial Intelligence (AI) has become a centrepiece of reshaping many aspects of individual life, society, and public and private sectors for the better \cite{gruetzemacher2022transformative,balahur2022data,maragno2023exploring}. Yet, concerns have been raised regarding the ethical implications and potential risks associated with the use of AI, such as biases in decision-making algorithms \cite{miron2021evaluating,araujo2020ai} and privacy issues \cite{zhang2021ethics,hupont2022landscape}. These concerns have prompted a global wave of trustworthy AI guidelines \cite{OECD,hlegtai} and some preliminary regulatory efforts aimed at mitigating the potential harms of AI, ensuring its development and use are aligned with safety standards and respect human rights and freedoms. The European Union's (EU) AI Act \cite{aiact}\footnote{For this work, we examined multiple AI Act mandates published since April 2021, in particular the agreed provisional text. However, the references to the AI Act within this paper shall be interpreted as references to the European Commission's proposal; as at the time of writing, the final content of the AI Act is not published in the official journal of the European Union.} 
is the first comprehensive legal framework on AI, through which AI systems are subjected to a set of regulatory obligations according to their risk level—that can be, from highest to lowest, \emph{unacceptable}, \emph{high-risk}, \emph{limited-risk}, or \emph{minimal-risk}. 

With the risk level being the yardstick for determining the regulatory requirements that AI systems need to satisfy, the AI Act is intent on promoting secure, trustworthy, and ethical use of AI, with a particular emphasis on management and mitigation of potential harms that high-risk AI systems may cause. The risk management system provisions, set in Art. 9, thereby play a pivotal role in the implementation of the Act. Effective AI risk management requires information regarding the AI system, its incorporating components, its context of use, and its potential risks to be maintained and communicated in the form of \emph{documented information}. While maintaining and sharing information regarding AI systems and their risks promote transparency and in turn trustworthiness \cite{hupont2023documentinghighrisk}, it is additionally a legal obligation for providers of high-risk AI systems to draw up technical documentation to demonstrate compliance with the Act's requirements (Art. 11). 
The elements of technical documentation, which includes risk management system,
are described in Annex IV at a high-level. However, to serve its purpose technical documentation needs to be extensive and detailed. Being limited to defining essential requirements, the AI Act relies on European \emph{harmonised standards}
for technical specifications to help with implementation and enforcement of its legal requirements, including technical documentation \cite{mazzini2023proposal,tartaro2023towards}. Notably, the European Commission's draft standardisation request \cite{EC2022request} refrains from calling for European standards in relation to data, model, or AI system documentation \cite{golpayegani2023high}. Consequently, the European Standardisation Organisations' work plan does not include standards specifically addressing these aspects \cite{soler2023analysis}. In parallel, the existing body of work on AI, model, and data documentation approaches pays little attention to documentation for legal compliance, fails to provide a strong connection between the documented information and responsible AI implications, and does not take into consideration aspects related to risk management \cite{hupont2023documentinghighrisk,heger2022understanding}.

Currently, there is a critical need for technical documentation to support and be in sync with the AI system development and usage practices. Moreover, with the involvement of several entities across the AI value chain, the documentation must be generated and maintained in a manner that ensures consistency and interoperability. This is also crucial for investigation of technical documentation by AI auditors and conformity assessment bodies for compliance checking and certification. 
 The motivation for this work is therefore to address this current need for consistent, uniform, and interoperable specifications to support effective implementation of the AI Act. This paper aims to address the current lack of unified technical documentation practices aligned with the AI Act, with a threefold contribution:
\begin{enumerate}
    \item we provide an in-depth analysis of the provisions of the EU AI Act in regard to documentation with a focus on \emph{technical documentation} and \emph{risk management system documentation} (Section \ref{AIA_documentation}); 
    \item we propose \textbf{\emph{AI Cards}}—a novel framework providing a human-readable overview of a given use of an AI system and its risks (Section \ref{AICards}); and 
    \item we present a machine-readable specification for AI Cards that enables generating, maintaining, and updating documentation in sync with AI development. It further supports querying and sharing information needed for tasks such as compliance checking, comparing multiple AI specifications for purposes such as AI procurement, and exchanging information across the value chain (Section \ref{semantic_representation}). 
    
\end{enumerate}
   
We demonstrate a practical application of AI Cards using an illustrative example of an AI-based proctoring system (Section \ref{use-case}), validate its usefulness through a survey (Section \ref{validation}), and discuss its benefits (Section \ref{benefits}).

\section{Related Work} \label{relatedwork}

\subsection{Previous Studies on the AI Act's Documentation and Risk Management Requirements}

Since the release of the AI Act's proposal by the European Commission in April 2021, the research community, international organisations, and industrial actors have been exploring the new avenues it opened by analysing the AI Act's contents for suggesting clarifications, identifying potential gaps, or giving critique. In this section, we refer to some existing studies on the provisions of the AI Act in regard to documentation and risk management. 

Panigutti et al. explore the role of the AI Act's transparency and documentation requirements in addressing the opacity of high-risk AI systems \cite{panigutti2023role}. Gyevnara et al. discuss compliance-oriented transparency required to satisfy the AI Act's requirements in regard to risk and quality management systems \cite{gyevnara2023get}. Schuett presents a comprehensive analysis of the AI Act's risk management provisions, without delving into the details of risk management system documentation 
\cite{schuett2023risk}. Soler et al.'s assessment of international standards in regard to addressing AI Act's requirements 
shows that these standards are insufficient to meet the provisions of the AI Act related to risk management \cite{soler2023analysis}.

In regard to technical documentation (Art. 11 and Annex IV), the most comprehensive studies, to the best of our knowledge, are the work of Hupont et al. \cite{hupont2023documentinghighrisk}, which identifies 20 information elements needed in documentation of AI systems and their constituting datasets, and the analysis of Annex IV provided by Golpayegani et al. \cite{golpayegani2022airo}, which identifies 50 information elements. Nevertheless, both studies remain short in being fully comprehensive in terms of covering both technical and risk management system documentation requirements laid down in Art. 11, Annex IV, and Art. 9. Building on our previous work, presented in \cite{golpayegani2022airo} and \cite{hupont2023documentinghighrisk}, this paper provides an in-depth analysis of technical documentation with a focus on risk management system documentation.

\subsection{Alignment of Existing Documentation Practices with the AI Act}

Compliance with the AI Act's documentation requirements requires guidelines, standards, tools, and formats to assist with generation, communication, and auditing of technical documentation. With transparency being widely recognised as a key factor in implementing trustworthy AI \cite{hlegtai}, several 
documentation frameworks proposed by the AI community, such as Datasheets for Datasets \cite{gebru2021datasheets} and Model Cards \cite{mitchell2019model}, have become \emph{de-facto} practices. As these documentation approaches are widely-adopted, a key question is the extent to which they could be leveraged for regulatory compliance tasks. 
This question is investigated by the studies mentioned in the following. Pistilli et al. discuss the potential of Model Cards as a compliance tool and anticipate its adoption—among other existing documentation practices originated from the AI community—for compliance with the AI Act's documentation obligations \cite {pistilli2023stronger}. 
The work by Hupont et al. \cite{hupont2023documentinghighrisk}, which investigates the 6 most widely-used AI and data documentation approaches for their alignment with the implementation of documentation provisions, concludes that AI Factsheets \cite{arnold2019factsheets} offers a higher overall degree of information coverage, followed by Model Cards and the AI Classification Framework proposed by the Organization for Economic Cooperation and Development (OECD) \cite{OECD}. The research also demonstrates that while data-related information elements are well-covered by most documentation approaches, particularly Datasheets for Datasets \cite{gebru2021datasheets}, the Dataset Nutrition Label \cite{holland2020dataset}, and the Accountability for Machine Learning framework \cite{hutchinson2021towards}, they still do not cover technical information needs related to AI systems. This finding is further strengthened in a follow-up comparative analysis of 36 AI system, model, and/or dataset documentation practices, which unravels the overall alignment of documentation practices with transparency requirements of the EU AI Act and other recent EU data and AI initiatives, and spots a gap in representing the information related to AI systems in its entirety and its context of use \cite{micheli2023landscape}. Interestingly, to date, the only documentation methodology conceived from design for the AI Act is \emph{Use Case Cards} \cite{hupont2023usecasecards}, which primarily focuses on documenting the intended use of AI systems to assess their risk level as per the AI Act, without mentioning technical information and risk management details.

\subsection{Machine-Readable Data, Model, and AI System Documentation} \label{sota-machine-readable}

While documentation approaches are acknowledged as instruments for improving transparency, and in turn enhancing trustworthiness, there has been little attention to the barriers in generation, maintenance, assessment, and exchange of conventional text-based documentation. Providing machine-readable specifications
is an idea taken up by some recent work to support adaptable and interoperable documentation. In the area of data documentation, Open Datasheets\footnote{\url{https://github.com/microsoft/opendatasheets-framework}}, proposed by Roman et al., provides a metadata framework for documenting open datasets in a machine-readable manner \cite{roman2023open}. For machine-understandable documentation of machine learning models, Model Card Report Ontology (MCRO)\footnote{\url{https://github.com/UTHealth-Ontology/MCRO}}, developed by Amith et al., offers the metadata for the content of Model Card reports \cite{amith2022semanticmodelcards}. Linked Model and Data Cards (LMDC) present a schema for integration of Model and Data Cards in a data space to provide a holistic view of a model or an AI service \cite{donald2023semantic}. The Realising Accountable Intelligent Systems (RAInS) ontology\footnote{\url{https://w3id.org/rains}}, created by Naja et al., models the information relevant to AI accountability traces \cite{naja2022kg}. 
 
In the context of documentation for regulatory compliance, the W3C Data Privacy Vocabulary (DPV) \cite{pandit2024data}\footnote{\url{https://w3id.org/dpv/}} is developed based on the requirements of the EU GDPR (General Data Protection Regulation) and has been applied for representing Data Protection Impact Assessment (DPIA) information \cite{pandit2022dpia}, documenting data breach reports \cite{pandit2023breach}, and representing Register of Processing Activities (ROPA) \cite{ryan2022dpcat}.  
In our previous work, we proposed the AI Risk Ontology (AIRO)\footnote{\url{https://w3id.org/airo}} for describing AI systems and their risks based on the AI Act and ISO 31000 family of standards \cite{golpayegani2022airo}. In this work, AIRO is employed to provide the basis for the machine-readable representation of the AI Cards. However, it is important to note that ISO 31000 family of standards is not aligned with the AI Act due to some fundamental differences \cite{soler2023analysis}. AIRO is used while awaiting future standards to be developed in response to the European Commission's standardisation request. Therefore, it needs to be updated with future publication of harmonised standards related to AI risk management.

\section{Provisions of the EU AI Act Regarding Documentation of AI Systems} 
\label{AIA_documentation}

\begin{figure}[ht]
    \centering
    \fbox{\includegraphics[width=\textwidth]{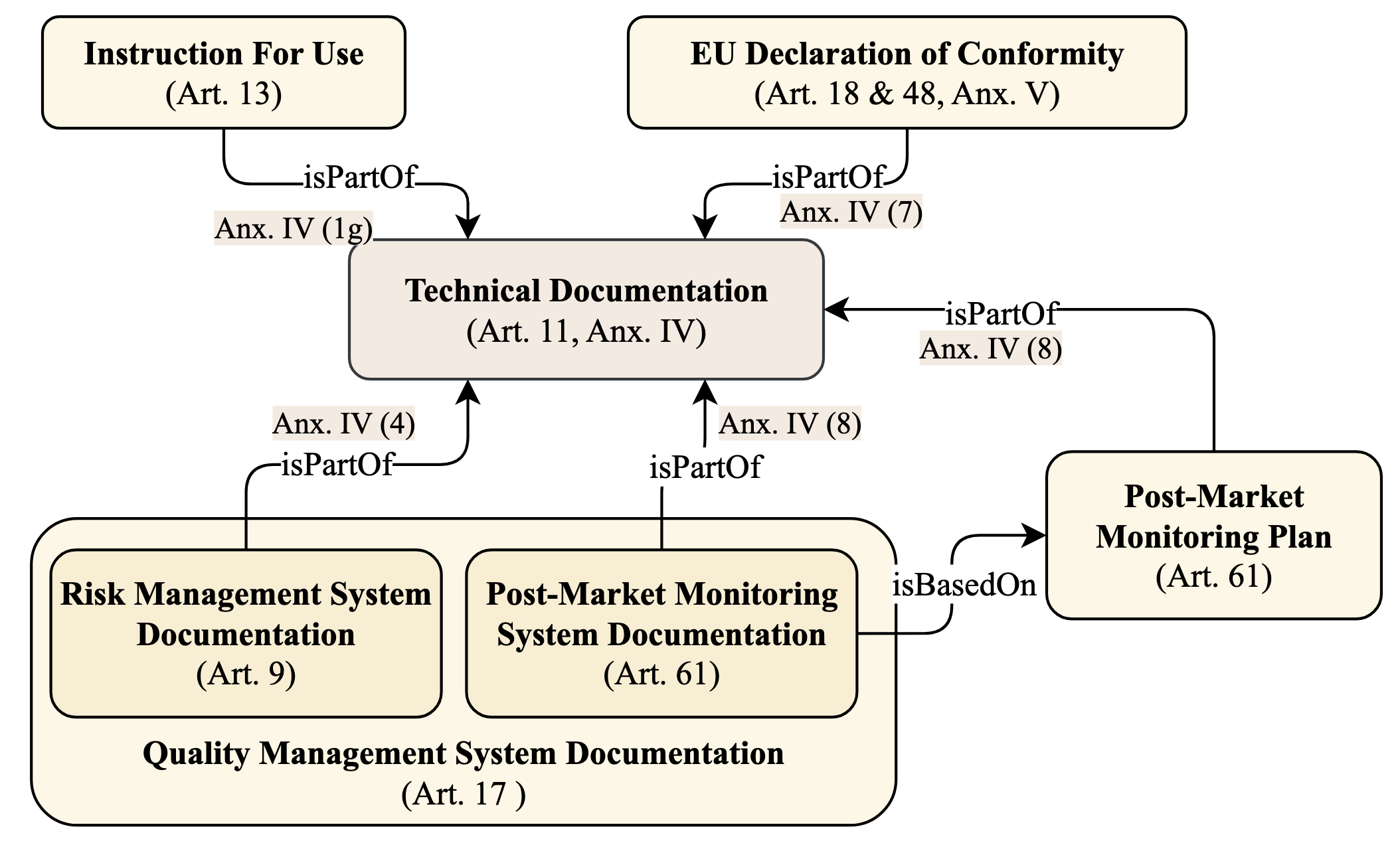}}
    \caption{High-risk AI systems'
    documentation requirements as per the AI Act.}
    \label{fig:aiact_key_documentation}
\end{figure}

Documentation of AI systems is highly tied to the principles of transparency and accountability \cite{hupont2023usecasecards,naja2022kg}, providing an appropriate degree of information to different stakeholders to enable several types of assessments. In a similar vein, documentation is essential for assessing legal compliance. Figure \ref{fig:aiact_key_documentation} shows the AI Act's documentation requirements for high-risk AI systems and illustrates how they are related to each other. Within the AI Act, \emph{technical documentation} and \emph{quality management system documentation} are key documents in the conformity assessment procedure, containing detailed information required for self- and third-party assessments. Given the central role of technical documentation and presence of related guidance on its content (Annex IV), this work focuses on identifying its concrete information elements. However, among its incorporating documents, our analysis expands on \emph{risk management system documentation} considering that the risk-centred nature of the AI Act makes the risk management system requirement (Art. 9) taking the lead in ensuring that the potential risks of high-risk AI systems are reduced to an acceptable level.  
In the following, we focus on identifying the information elements required to be featured in technical (Section \ref{technical_documentation}) and risk management (Section \ref{risk_management_documentation}) documentation.

\subsection{Article 11 and Annex IV - Technical Documentation} \label{technical_documentation}

As discussed earlier, technical documentation is a key artefact for demonstrating compliance with the AI Act. According to Art. 11, it is the \emph{high-risk AI provider} who should draw up the technical documentation \emph{ex-ante} and keep it \emph{updated}. From this obligation, three critical challenges arise: (i) when development of AI systems includes integrating third-party components into the system, e.g. training data and pre-trained models, high-risk AI providers need to dispense details regarding these components in technical documentation, e.g. information about how data is labelled (Annex IV(2)(d)). Thus, generating technical documentation to fulfill the requirements of the AI Act can become a collective activity, which requires communication and exchange of often sensitive and confidential information. This opens up a myriad of questions concerning trust, accountability, and liability, that lie beyond the scope of this discussion. (ii) Real-world testing environments and sandboxes assist in determining what might be difficult to ascertain ex-ante. However, specificities of the \emph{context of use}, which significantly influence AI risks and impacts, might not be revealed until the system's deployment. This suggests a need for ex-ante involvement of potential users and ex-post feedback loops. (iii) Finally, keeping technical documentation up-to-date and re-assessing legal compliance
are hard-to-manage tasks. A key question here is ``what changes do trigger the update process?''. A conspicuous, albeit not comprehensive, answer is \emph{substantial modifications}—changes that affect the system's compliance with the Act or its intended purpose (Art. 3(23)). However, lack of clarity regarding what modifications are deemed as substantial adds to the complexities of this challenge.

To help with the generation and auditing of technical documentation, the minimum set of its incorporating information elements is outlined in Annex IV, which is subject to the
European Commission's potential amendments (Art. 11(3)). Annex IV serves as a primary template wherein information elements are described with varying degrees of detail, with the majority articulated at a high-level. This implies a need for harmonised standards to support the implementation of Art. 11. However, as mentioned earlier, to the best of our knowledge there is no ongoing standardisation activities to address this need. Aiming to provide clarification on the content of technical documentation, we conducted an in-depth analysis of Annex IV, with a particular emphasis on the risk management system. 
The analysis is an initial step in establishing a collective understanding of the technical documentation's content and could inspire the development of prospective European standards, guidelines, and templates.

\subsection{Article 9 - Risk Management System Documentation} \label{risk_management_documentation}

 The AI Act will rely on harmonised standards
 to operationalise legal requirements at the technical level, and risk management system will be no exception. At the time of writing, however, European Standardisation Organisations are still working on deliverables planned in response to the European Commission's standardisation request. Furthermore, existing ISO/IEC standards on AI risk management and AI management system, namely ISO/IEC 23894\footnote{\url{https://www.iso.org/standard/77304.html}} and ISO/IEC 42001\footnote{\url{https://www.iso.org/standard/81230.html}}, 
 have been found insufficient in meeting the requirements of the AI Act \cite{soler2023analysis}.

 Indeed, within the current AI standardisation realm, the two main international standards covering AI risk management aspects that have been analysed in regard to AI risk management
 in the technical report published by the European Commission's Joint Research Centre \cite{soler2023analysis} are: 
 (i) \textbf{ISO/IEC 42001 ``Information technology — Artificial intelligence — Management system''} that offers a framework to assist with implementation of AI management systems, and (ii) \textbf{ISO/IEC 23894 ``Artificial intelligence — Guidance on risk management''}, which offers practical guidance on managing risks, albeit defined as the effect of uncertainty, without relation to potential harms to individuals. 
 Due to this and other reasons, the requirements defined in these standards are not useful to comply with the provisions of the AI Act in regard to risk management \cite{soler2023analysis}, however they may be useful to identify relevant technical information elements for the development of AI Cards while we await harmonised standards for the AI Act.
 In this section, we elaborate on information elements for the documentation of an AI risk management system by obtaining insights from these two standards.

\begin{figure}[ht]
    \centering
    \fbox{\includegraphics[width=\textwidth]{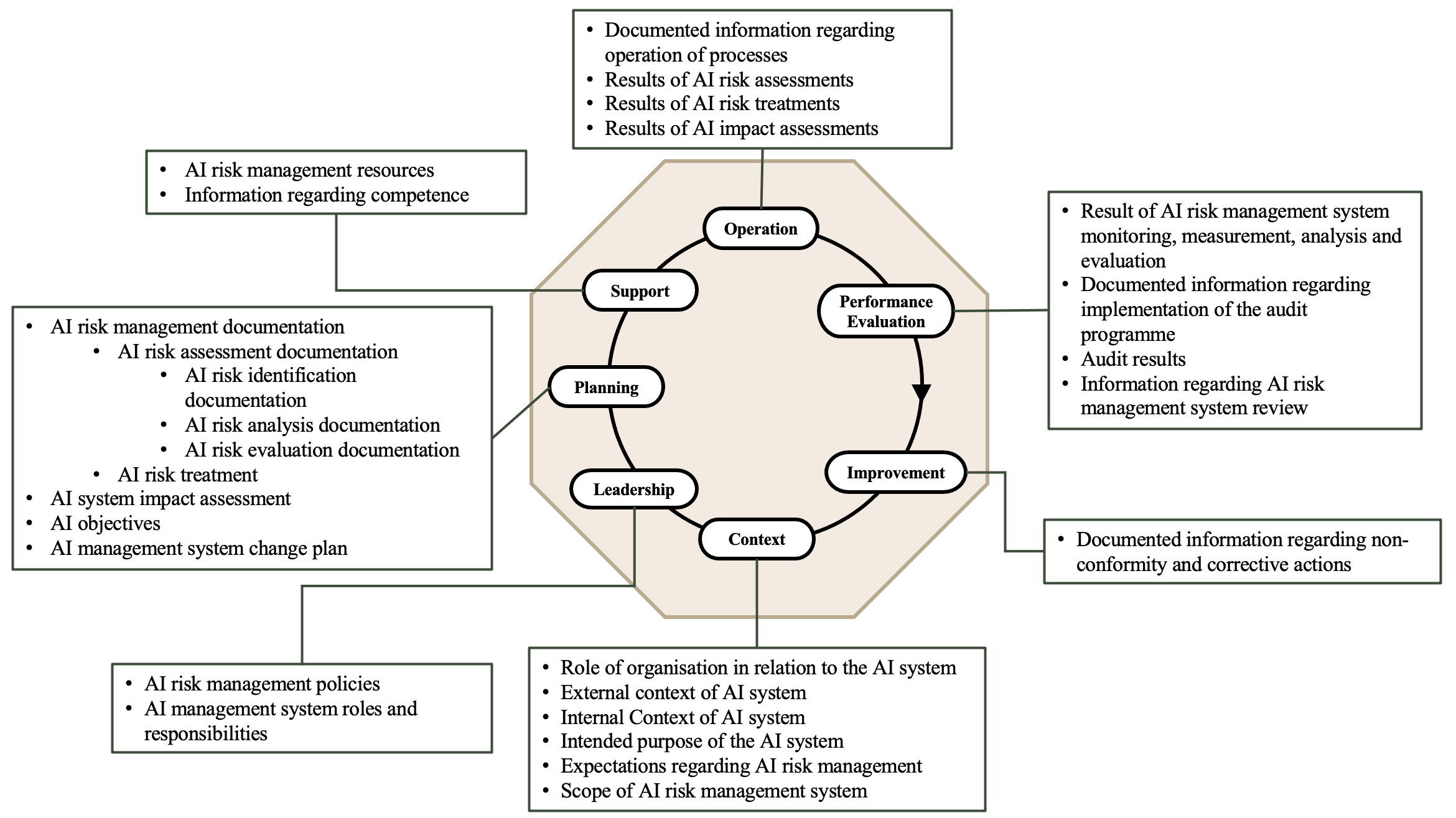}}
    \caption{Information elements identified for AI risk management system documentation (note they are inspired by ISO/IEC 42001 and 23894, and therefore will be updated according to the AI Act's future official harmonised standards).} 
    \label{fig:AIRMS}
\end{figure}

 We used the overall structure of AI management systems, which follows the ISO/IEC's \emph{harmonised structure} for management system standards\footnote{\url{https://www.iso.org/sites/directives/current/consolidated/index.html}}, as a resource for extracting the key organisational activities within an AI risk management system. For each activity, documentation needs, in terms of information elements, were identified from both ISO/IEC 42001 and 23894 (for a summary of the analysis see Figure \ref{fig:AIRMS}). These information elements can be categorised into four overall groups: 
\begin{itemize}
    \item Information about the \emph{context of the AI system and the organisation}, for example, the AI system's intended purpose and the role of the organisation in relation to the system.
    \item Details of the \emph{risk management system} in place, e.g. the policies, responsibilities, and resources required for implementation of the risk management system itself. This category of information is relevant to the ISO/IEC 23894's AI risk management framework, whose intention is to help with AI governance and integration of AI risk management activities into an organisation's existing processes.
    \item Documentation of \emph{risk management processes} across different phases: planning (ex-ante), operation (ongoing), and post-operation (ex-post). 
    \item \emph{Results of AI risk management}, which can be represented in artefacts produced throughout the risk management process, e.g. risk assessment documentation that lists AI risks identified, their likelihood, severity, sources, consequences, and impacts.  
\end{itemize}

\section{AI Cards} \label{AICards}

Extensiveness of technical documentation and confidentiality of its content may hinder collaboration and communication with AI stakeholders—many of whom are not necessarily technical AI experts. To address this challenge, we propose the \textbf{AI Cards framework}, as a structured information sheet, which offers an overview of a use of an AI system and its related trustworthy AI concerns by inclusion of crucial information about technical specification, context of use, risks, and compliance. With an intuitive representation, the AI Cards framework introduces a simple, transparent, and comprehensible summary card laying out a holistic picture of a specific AI use case and its risks without disclosing sensitive information. The framework encompasses machine-readable specifications that enhance interoperability, enable automation, and support extensibility. 

\subsection{Development Process} 

 In shaping the framework, we first specified its requirements, which are summarised in Table \ref{tab:design}. Based on the defined scope, we re-examined our analysis of Annex IV. The information elements within the scope were considered as candidates for inclusion in the AI Cards. However, many of them were 
 too detailed to be represented in a summary card. Therefore, we propose condensed views for these information elements through visual aids. 

The AI Cards framework was defined and iteratively refined in consultation with researchers involved in digital policymaking within the European Commission and further validated by conducting an online anonymous survey with law and technology researchers (see Section \ref{validation} for more details).

\begin{table}[]
    \centering
    \caption{AI Cards framework requirements}
    \begin{tabularx}{\textwidth}{p{3cm} X}
    \hline
    Purpose & Providing an overview of technical documentation that effectively conveys key information 
    \\
    \hline
    Scope &   
        In the scope: AI system, as a whole, its context of use, and information relevant to trustworthy AI concerns including risk management system.
        
        Out of the scope: organisational processes, details of management systems, and documentation of AI components, including data and model. The framework is horizontal and does not take sector-specific nuances into account.
      \\
    \hline
    Key audience & AI users, end-users, subjects, providers, developers, auditors, and policymakers\\
    \hline
    Representation formats & An easy-to-understand visual human-readable representation accompanied with a machine-readable specification\\
    \hline
    
    \end{tabularx}
    
    \label{tab:design}
\end{table}

\subsection{Information Elements}
 
\begin{figure}[!ht]
    \centering
    \fbox{\includegraphics[width=0.8\textwidth]{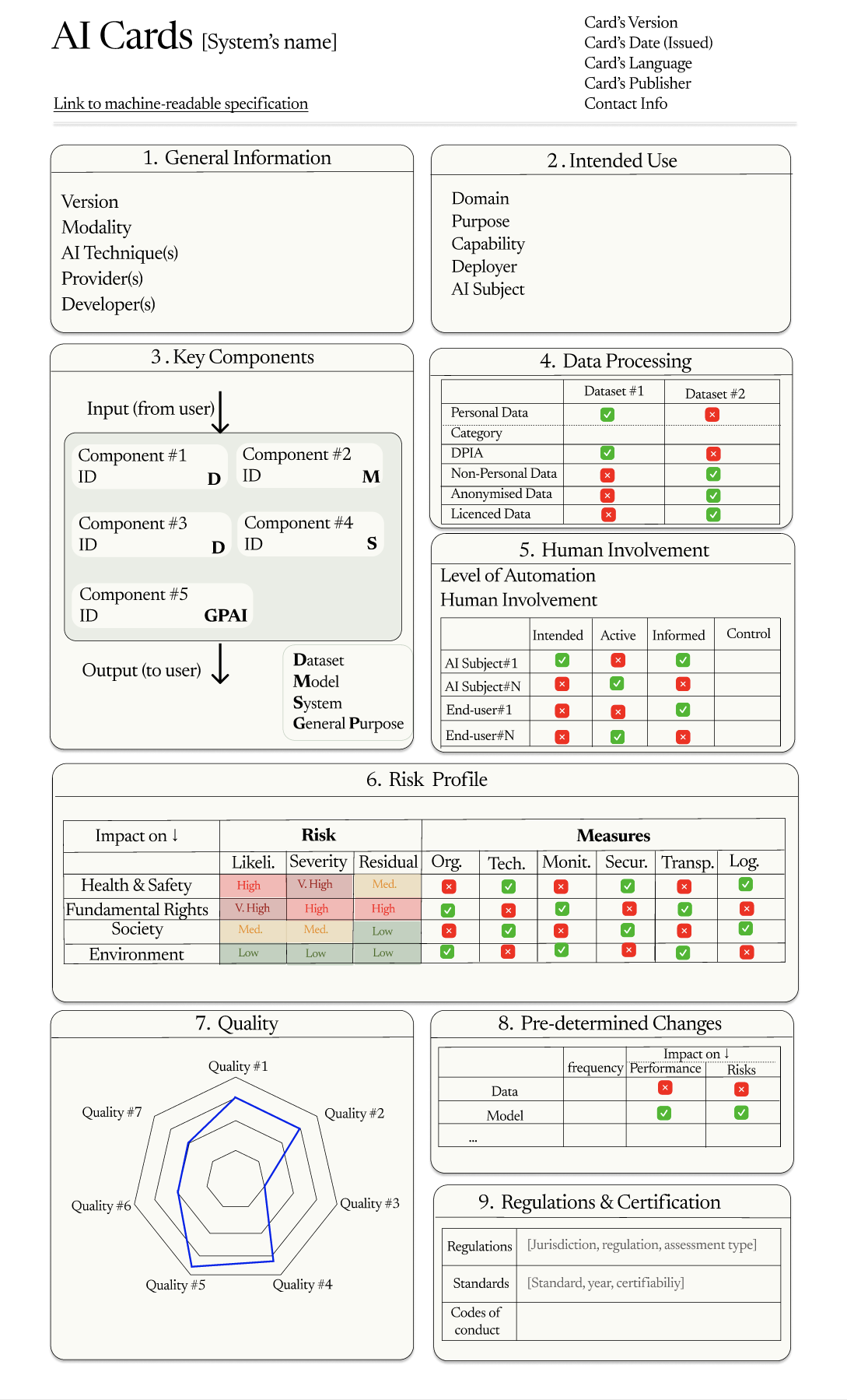}}
    \caption{Visual representation of AI Cards.}
    \label{fig:AICards}
\end{figure}

 Figure \ref{fig:AICards} shows a visualisation of the AI Card which condenses the information elements in 9 sections. To enable representing AI Cards in a time series and to allow version control required to reflect the evolving nature of AI systems, the Card's metadata includes essential information describing its \textbf{version}, \textbf{issuance date}, \textbf{language}, \textbf{publisher}, and \textbf{contact} information. Further, the \textbf{URL of the machine-readable specification} is included.

\subsubsection{(1) General Information}
This section provides the key information about the AI system including its \textbf{version}, \textbf{modality}, e.g. standalone software or safety component of a product, main \textbf{AI techniques} used, \textbf{provider(s)}, and \textbf{developer(s)}. 

\subsubsection{(2) Intended Use} 
The AI Act put a considerable emphasis on intended use of the system, given its profound effect 
 on risks and impacts. 
For describing the intended use, we propose using the combination of the 5 concepts identified in \cite{golpayegani2023high}, which are:
 \begin{itemize}
     \item the \textbf{domain} in which the system is intended to be used within,
     \item the \textbf{purpose}, i.e. end goal, of using the system in a specific context,
     \item the \textbf{AI capability} that enables realisation of the purpose,
     \item the \textbf{AI deployer} which is the entity using the AI system, 
     \item the \textbf{AI subject} which is the entity subjected to the outputs, e.g. decisions, generated by the AI system.
 \end{itemize}
  
Noteworthy, the subsequent sections heavily rely of the information represented in this section, considering that they should be defined in the view of the intended use.

\subsubsection{(3) Key Components} 
Many AI systems are built through integration of multiple AI models, datasets, general-purpose AI systems, and other software elements, each of which has an effect on the system's behaviour and in turn its risks \cite{micheli2023landscape}. This section provides the system's \emph{high-level architecture} in terms of incorporating \textbf{components}. For each component, its \textbf{name}, \textbf{version}, and \textbf{link to documentation or ID} are presented. For detailed information about components, we rely on their documentation provided presumably by the components' providers in the form of \textbf{information sheets}, e.g. Datasheets, Model Cards, and AI Factsheets. 

\subsubsection{(4) Data Processing} 
AI systems that do not process data, if they exist at all, are rare. Within the EU digital acquis, the GDPR \cite{gdpr}, which protects the fundamental right to privacy by regulating \emph{personal data}, is applicable to those AI systems that process natural persons' data. Therefore, having knowledge
of whether a given use of an AI system involves processing of personal data is crucial to correctly interpret the resulting legal compliance obligations. 
This section specifies inclusion of processing \textbf{personal data} and shows the \textbf{category} of the data, e.g. biometric data, and represents whether data protection impact assessment \textbf{(DPIA)} is conducted. It further specifies inclusion of \textbf{non-personal}, \textbf{anonymised}, and \textbf{licenced} data.

\subsubsection{(5) Human Involvement} 
 Involvement can take different forms depending on the phase of AI development, the role of human actors, and the system's \textbf{level of automation}—which has a range from fully autonomous to fully human-controlled according to ISO/IEC 22989\footnote{\url{https://www.iso.org/standard/74296.html}}. The level of automation also has a substantial effect on the safeguards, including human oversight measures, required for controlling AI risks. This section provides an overview of involvement of two specific human actors: \textbf{AI end-users}, who use the AI system's output and \textbf{AI subjects}, who are subjected to the outputs of the system. For these actors, we look into the following aspects of involvement:

\begin{itemize}
    \item \textbf{Intended involvement}: represents whether the involvement of a specific actor is as intended. An example of an \emph{intended} AI subject in an AI-based proctoring system is a student sitting an online test. In this case, other persons present in the room are \emph{unintended} AI subjects.

    \item \textbf{Active involvement}: shows whether a specific actor actively interacts with the AI system. 
    
    \item \textbf{Informed involvement}: represents whether a specific actor was informed that an AI system is in place; for example, 
    in cases where a decision affecting a person's education is made using AI-based solutions.
    
    \item \textbf{Control over AI outputs}: shows the extent to which AI subjects and end-users have control over AI outputs, in particular decisions made by the AI system and their impacts. Inspired by OECD's modes of operationality \cite{OECD}, we consider the following six levels of control: 
    \begin{itemize}
        \item An AI subject/end-user can opt in the system’s output.
        \item An AI subject/end-user can opt out of the system’s output.
        \item An AI subject/end-user can challenge the system’s output.
        \item An AI subject/end-user can correct the system’s output.
        \item An AI subject/end-user can reverse the system’s output ex-post.
        \item An AI subject/end-user cannot opt out of the system’s output.  
        \end{itemize}
\end{itemize}

\subsubsection{(6) Risk Profile}
 This section provides a high-level summary of risk management results, which includes an overview of \textbf{likelihood}, \textbf{severity} and \textbf{residual risk} associated with risks that have impact on areas of \textbf{health and safety}, \textbf{fundamental rights}, \textbf{society}, and \textbf{environment}. Further, this section shows whether any technical, monitoring (human oversight), cybersecurity, transparency, and logging measures applied to control the risks. 

\subsubsection{(7) Quality} 
With many AI incidents rooted in poor quality, ensuring that the system is of high quality by participating in AI regulatory sandboxes, benchmarking, and performing tests is necessary. We propose illustrating key \textbf{AI system qualities} using a radar chart. We recommend including \textbf{Accuracy}, \textbf{robustness}, and \textbf{cybersecurity}, which are the key qualities explicitly mentioned in the Act (Art. 15). Further, the relevant \emph{product quality} and \emph{quality in use}, described by ISO/IEC 25059 on AI SQuaRE (Systems and software Quality Requirements and Evaluation)\footnote{\url{https://www.iso.org/standard/80655.html}} can be considered to be included in the quality section.

\subsubsection{(8) Pre-determined Changes} 
This section provides a list of \textbf{pre-determined changes} to the system and its performance in terms of \textbf{subject} and \textbf{frequency} of change as well as the potential \textbf{impacts of change on performance and risks}.      
 
\subsubsection{(9) Regulations \& Certification}
This section lists the main digital \textbf{regulations} the AI system is compliant with, key \textbf{standards} to which the system or the provider(s) conform, 
and \textbf{codes of conduct} followed in development or use of the AI system.

\subsection{Machine-Readable Specification}  \label{semantic_representation}

The visual representation assists stakeholders in gaining an understanding of an AI system, its context, and the associated trustworthy AI concerns without delving into the extensive details of technical documentation. However, these are not sufficient to support some of the desirable features of documentation, including search and tracking capabilities, 
conducting meta-analysis, comparing multiple AI systems, and automating generation and update of documentation \cite{heger2022understanding}. To include these features, the AI Cards framework supports machine-readable representation of information by leveraging the standards, methods, and tools provided by the World Wide Web Consortium (W3C)\footnote{\url{https://www.w3.org/}}, motivated by the body of work discussed in Section \ref{sota-machine-readable} as well as the rise in uptake of open data formats for documentation, reporting, and sharing information by the AI community, e.g. HuggingFace’s use of JSON Model Cards\footnote{\url{https://huggingface.co/docs/hub/model-cards}}, and by the EU, e.g. machine-readable regulatory reporting \cite{eu2022mrer}, DCAT-AP open data portals\footnote{\url{https://op.europa.eu/en/web/eu-vocabularies/dcat-ap}}, and EU vocabularies and ontologies\footnote{\url{https://op.europa.eu/en/web/eu-vocabularies/controlled-vocabularies}}.
The machine-readable representation, which relies on Semantic Web technologies, fosters openness and interoperability—both essential for exchanging information across the AI value chain. This representation is extensible and therefore enables accommodating sector-specific information requirements and allows adaptation to highly-anticipated guidelines from authorities including the AI office, the potential amendments to the Act via delegated Acts, and case law. Grounded on formal logic, it also assists in ensuring that the information is complete, correct, and verifiable.

\begin{figure}[ht]
    \centering
    \fbox{\includegraphics[width=\textwidth]{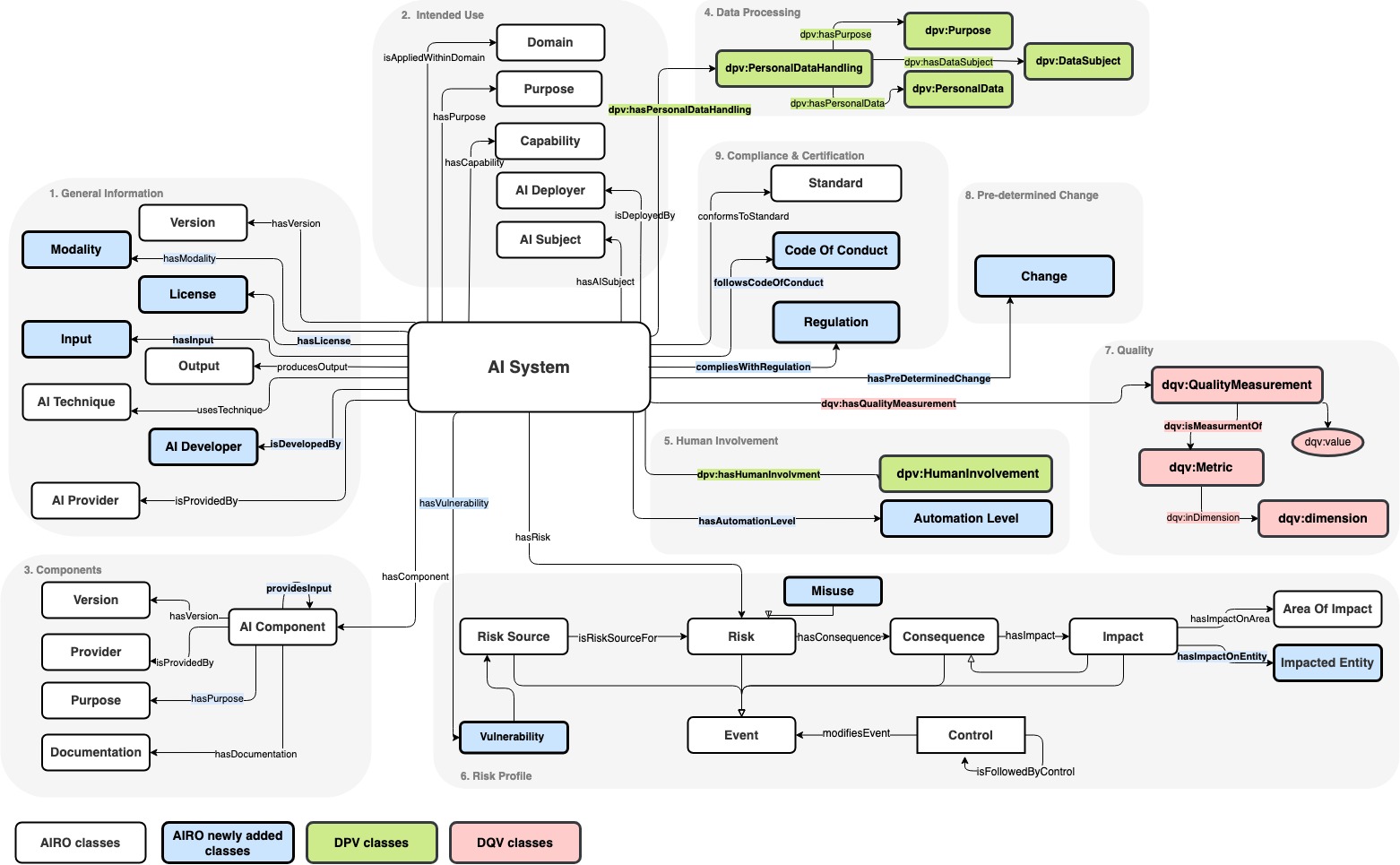}}
    \caption{An extension of AIRO for AI Cards.}
    \label{fig:semantic_model}
\end{figure}

For provision of machine-readable specifications for AI Cards, an \emph{ontology} is needed to provide a common semantic model that explicitly represents terms, their definitions, and the semantic relations between them, ensuring consistency and interoperability. For this, building upon our previous work \cite{golpayegani2022airo}, we further extend AIRO to support modelling information elements featured within the AI Cards. This extension borrows concepts and relations from Data Quality Vocabulary (DQV)\footnote{\url{https://www.w3.org/ns/dqv}} \cite{albertoni2021dqv} for expressing AI quality and Data Privacy Vocabulary (DPV)\footnote{\url{https://w3id.org/dpv/}} \cite{pandit2024data} for modelling personal data processing. Figure \ref{fig:semantic_model} depicts key classes and relations required for representing information featured in each section of the AI Card.

With the ontology providing the schema, the information incorporated in the AI Cards can be represented as an RDF graph. This opens the door to leveraging the power of Semantic Web standards and technologies across a variety of tasks, among them are: 
\begin{itemize}
    \item Querying to retrieve information about an AI system and information related to demonstration and investigation of legal compliance, for example verifying presence of measures to address potential AI impacts on the fundamental right to non-discrimination in support of compliance with Art. 9 of the AI Act. The information retrieval is implemented using \texttt{SELECT} queries in the SPARQL\footnote{\url{https://www.w3.org/TR/sparql11-query/}}, which the W3C standardised query language, as shown in \cite{golpayegani2022airo}. 
    
    \item Updating information about an AI application, for example when a mitigation measure is no longer effective and is replaced with a new measure, which is implemented using \texttt{DELETE/INSERT} SPARQL queries. 
    
    \item Integrating information provided in documentation of incorporating components (demonstrated in the ``Key Componets'' section of the AI Card), for example a Model Card that documents a system's incorporating model, when provided in standardised linked data formats such as JSON-LD or RDF.      

    \item Reasoning that might support conformity assessment and legal compliance tasks, for example defining semantic rules to suggest the regulatory risk profile associated with an AI application. Semantic rule-checking using the Shapes Constraint Language (SHACL)\footnote{\url{https://www.w3.org/TR/shacl/}}, which is a W3C standardised language, is valuable in validation of information against a set of rules (see prior work on determining high-risk AI applications using SHACL \cite{golpayegani2023high}). 
    
    \item Expressing intended use policies in agreements between different parties in the AI value chain, for example agreements between AI providers and deployers describing conditions of using or modifying an AI application. In this, the Open Digital Rights Language (ODRL)\footnote{\url{https://www.w3.org/TR/odrl-model/}}—the W3C recommended language for representing policies—can be used for describing intended and precluded uses as \emph{permission} and \emph{prohibition} statements respectively.
\end{itemize}

 \section{Proof-of-Concept: AI Cards for an AI-Based Proctoring System}  \label{use-case}

 To demonstrate the potential scalability, and applicability of the proposed framework, and inspired by the use cases described in \cite{panigutti2023role} and \cite{hupont2023usecasecards}, we take an AI-based student proctoring tool called \emph{Proctify} as an illustrative proof-of-concept. It is important to note that this system is merely selected to demonstrate the applicability of the AI Cards framework in real-world cases. The AI Act generally considers such proctoring systems as \emph{high-risk} in the education domain when used for monitoring and detecting suspicious behaviour of students during tests \cite{hupont2022landscape}.
 
 Proctify is intended to detect suspicious behaviour during online exams by analysing facial behaviour from a student's facial video captured throughout the exam using a webcam. Prior to this, students have explicitly consented to be recorded during the exam and informed that they must be alone in the room. The system incorporates a graphic interface displaying an analysis of the student's face including the head pose, gaze direction, and face landmarks' positions. This extracted information is then provided as an input to \emph{SusBehavedModel}, which has been trained in-house by the system's provider using \emph{SusBehavedDataset}, to determine whether the student is displaying suspicious behaviour, e.g. looking away from the screen, leaving the room, or a third person detected in the room. Detection of suspicious behaviour raises an alarm in the interface to inform and let the human oversight actors, e.g. human instructors, take appropriate actions, e.g. communicating with the student. 
 
\begin{figure}
    \centering    \fbox{\includegraphics[width=\textwidth]{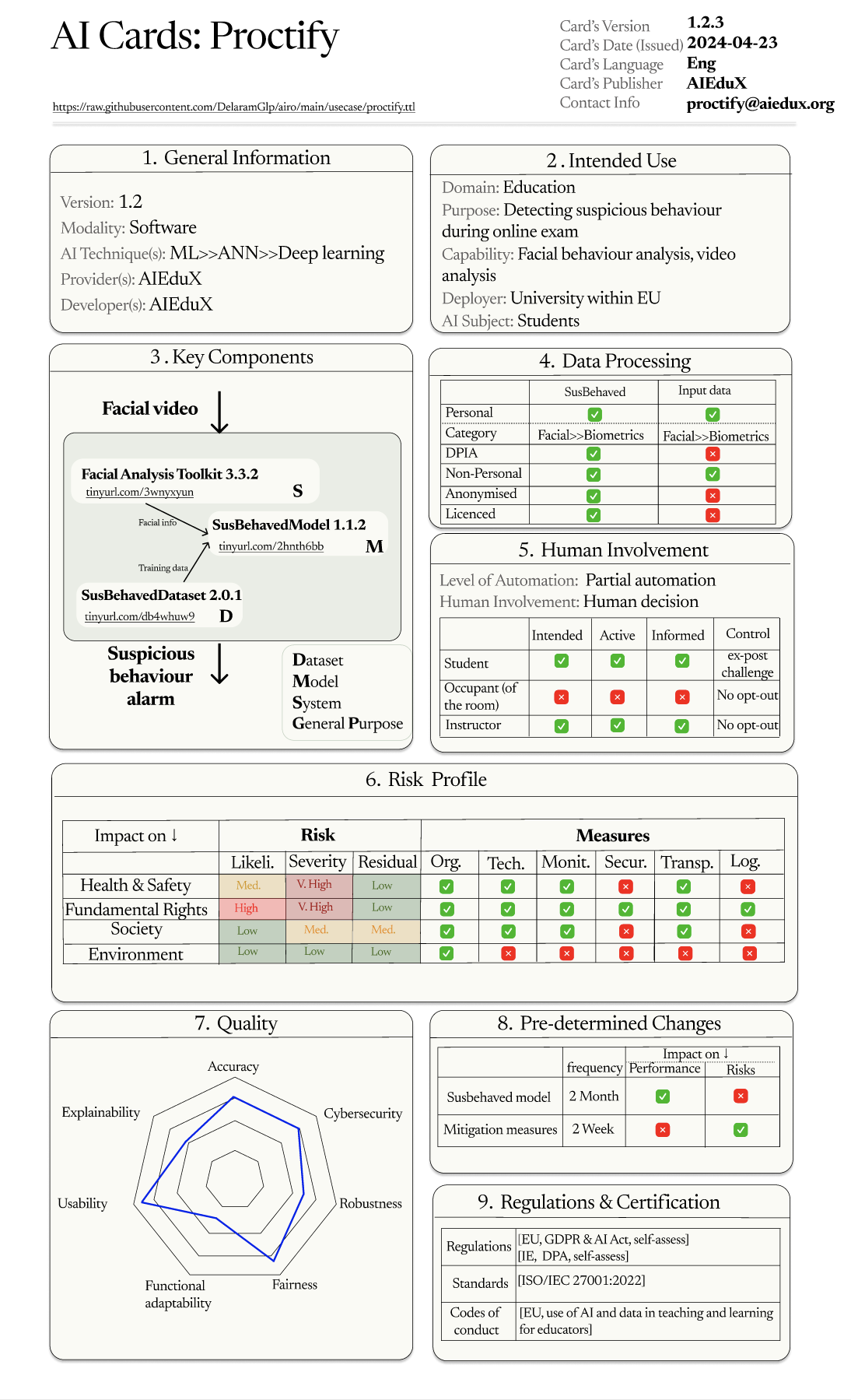}}
    \caption{An example of AI Card for an AI-based proctoring system.}
    \label{fig:AICard_proctify}
\end{figure}

 Throughout the risk management process, the risks and impacts of the system are identified and assessed by the provider, including the following: the system may have lower accuracy for students with darker skin tones and a higher rate of false-positive alarms for students wearing glasses. Further, false-negatives and false-positives are more frequent for students with health issues or disabilities that affect their facial behaviour. There is also a chance of over-reliance of human instructors on the system's output (automation bias). These events have the potential to negatively impact students' \emph{mental health}, \emph{future career}, and their rights to \emph{dignity} and \emph{non-discrimination}. 
 Some of the measures applied to address the system's risks and impacts are: ensuring the dataset is representative and diverse in demographic terms, conducting rigours and frequent testing of accuracy, assigning expert human proctors, and creating clear protocols to act upon when an alarm is raised.
 The Proctify's AI Card is visualised in Figure \ref{fig:AICard_proctify} and its machine-readable specification is available online\footnote{\url{https://github.com/DelaramGlp/airo/blob/main/usecase/proctify.ttl}}. 
 As shown in the figure, the visual representation of the AI Card provides a summary of risk management information without disclosing the details. However, the machine-readable specification is capable of modelling the details.%

\section{Validation} \label{validation}
As mentioned earlier, to identify essential information elements of the AI Cards framework, we consulted experts involved in EU digital policymaking, in particular the AI Act, whose backgrounds are in AI transparency and documentation, explainability, cybersecurity, standardisation, and Semantic Web. Throughout the development process, our analysis of the AI Act's provisions, the visual representation of AI Cards, and its machine-readable specification were discussed to ensure alignment with EU digital policies, adoption of correct terminology, and suitability of the AI Cards framework for addressing common concerns of AI stakeholders. After reaching a solid structure for the AI Cards, an online anonymous survey has been conducted to assess the usefulness of the AI Cards framework.

\subsubsection{Survey Design} 
In addition to questions regarding the respondent's background, the survey includes questions about usefulness of (i) (only) the visual representation, (ii) (only) the machine-readable representation, and (iii) the overall AI Cards framework (both human- and machine-readable representations). In the latter, to assess how stakeholders envisage usability of the AI Card the System Usability Scale (SUS) \cite{brooke1996sus} is used. As the SUS is originally designed for assessing perceived usability, minor wording changes were applied to fit it with the purpose of AI Cards' assessment and to provide more clarity. Table \ref{tab:sus} shows the list of SUS-based questions for assessing the AI Cards' usability, each answered on a 5-point Likert scale ranging from 1 (strongly disagree) to 5 (strongly agree).

\subsubsection{Recruitment}

In the first phase, a student cohort of participants (N=23) were recruited from the Data Governance module in Dublin City University attended by students enrolled in Masters of Art (MA) in Data Protection and Privacy (MDPP) and the European Master in Law, Data and Artificial Intelligence (EMILDAI). It is important to highlight that, because of the nature of this module, some of the participants (30 percent) have notable positions in public organisations, industry, and NGOs. It should be noted that a second phase of evaluation with policymakers and industry stakeholders is currently underway.

\subsubsection{Preliminaries}
Prior to carrying out the survey, it was reviewed and approved by the ethics committee within the school of computer science and statistics of Trinity College Dublin. To start, in a preliminary interactive session, information about the AI Act and the AI Cards was presented to participants and they were asked to sign the informed consent form. Then, they could start completing the survey. It should be noted that the participation in the survey was optional.

\subsubsection{Findings}
23 responses were collected in this first phase of the evaluation from the student cohort. Regarding the usefulness of the visual representation, inclusion of the following information was suggested in the survey results:
\begin{itemize}
    \item information related to the registry process,
    \item link to the GDPR compliance including summary of the requirements for data processing and retention period of data if personal data processed, 
    \item limitations of the system,
    \item information regarding tests including tests, bugs, issues, and date of testing,
    \item a legend or key to explain the abbreviations.
    
\end{itemize}

In the context of implementation and enforcement of the AI Act, the participants agreed on the usefulness of the human-readable representation in compliance checking, creating technical and risk management system documentation, and creating guidelines. The participants further referred to the following potential uses of the AI Cards: \emph{ensuring traceability and transparency}, \emph{monitoring the system}, \emph{reviewing periods for pre-determined changes}, \emph{raising awareness}, and \emph{explaining the core concepts to stakeholders with different levels of AI literacy and technical knowledge}. In regard to usefulness of AI Cards as a tool to facilitate exchange of information regarding an AI system and its risks, while all participants acknowledged its usefulness to some extent for communication within the AI development ecosystem and with authorities, a minority (4 out of 23) did not perceive it valuable for communication with the public. 

Regarding machine-readable specifications, there was general consensus on its usefulness for automated tools, establishing a common language, and structuring the EU registry of AI systems. Further, the participants indicated that the machine-readable specifications could assist in \emph{determining the legal risk category}, \emph{providing a better understanding of models}, and \emph{building models with better compliance capabilities}.

Regarding the potential target users of the framework, one participant referred to 
\emph{``AI Brokers for comparing AI Systems - would require more metrics but a good start for comparison''} and another participant mentioned \emph{``Members of the public, in fostering awareness and understanding of AI systems''}. The results of SUS usability test are shown in Table \ref{tab:sus}, where due to the mixed tone of the items a higher score for odd-numbered items and a lower score for even-numbered items are desired. The average SUS score, calculated using the formula proposed in \cite{lewis2018sus}, is 66.30.

\begin{table}[!h]
    \centering
    \caption{SUS questions used for evaluating usability of AI Cards}
    \begin{tabularx}{\textwidth}{p{0.5cm} X  p{1.5cm}}
    \hline
     No. & Question & Score (mean) \\
     \hline
     1 & I think that AI stakeholders would like to use the AI Cards framework frequently. & 4.21 \\
     \hline
     2 & I find the AI Cards framework unnecessarily complex. & 2.13 \\
     \hline
     3 & I think the AI Cards framework is easy to use. & 3.96 \\
     \hline
     4 & I think that AI stakeholders would need the support of a technical person to be able to use the AI Cards framework. & 3.35  \\
     \hline
     5 & I find the various aspects (i.e. information elements and human- and machine-readable formats) in the AI Cards framework are well integrated. & 4.22 \\
     \hline
     6 & I think there is too much inconsistency in the AI Cards framework. & 1.83 \\
     \hline
     7 & I would imagine that most people would learn to use the AI Cards framework very quickly. & 3.91 \\
     \hline
     8 & I find the AI Cards framework very cumbersome to use. & 2.26 \\
     \hline
     9 & I feel very confident using the AI Cards framework. & 3.43 \\
     \hline
     10 & I need to learn a lot of things before I could get going with the AI Cards framework. & 3.30 \\
     \hline
    \end{tabularx}
         
    \label{tab:sus}
\end{table}

Another key aspect in evaluation of effectiveness and usefulness of an AI documentation framework is the extent of its adoption by the AI community, as evidenced in widely-adopted documentation approaches, e.g. Datasheets \cite{gebru2021datasheets}, Model Cards \cite{mitchell2019model}, and Factsheets \cite{arnold2019factsheets}. Given that the AI Act was only adopted recently and the final text has not been published yet, the extent of AI Cards’ adoption will need to be assessed over time. As indicated by one the participants, \emph{``The real world application would be interesting to see and may require changing based on different Member State authorities requiring other information be made available for documentation purposes''}.

\section{Benefits and Potential Applications} \label{benefits}

The AI Cards framework distinguishes itself from existing documentation approaches through: 
\begin{itemize}
    \item Its \textbf{alignment with the EU AI Act's provisions} in regard to technical and 
    risk management system documentation. It should be noted that though the framework is designed based on the requirements of the EU AI Act, its modular and semantic nature makes it scalable to be applied 
    to document a wide range of AI applications, regardless of the jurisdiction they are used in.
    \item Its dual approach towards information representation, which makes the framework \textbf{accessible} to 
    both humans and machines, facilitates 
    communication, promotes \textbf{interoperability}, and enables automation.  
    \item Its holistic \textbf{AI
    system-focused} approach, which allows inclusion of 
    information regarding the context of use, risk management, and compliance in addition to technical specifications. 
    While most existing documentation practices address potential risks or ethical issues at a very high and unstructured level \cite{hupont2023documentinghighrisk}, an AI Card draws a picture of the \textbf{risk management system} that is already established by illustrating an overview of identified risks and applied mitigation measures.  
    \item Its \textbf{modular and future-proof} design that enables reconfiguration and extension of the framework to address documentation requirements 
    arising from forthcoming AI regulations, policies, and standards.
\end{itemize}

AI Cards' machine-readable specification adds an extra level of potential for application. In terms of documentation management, this specification facilitates frequent modification, 
version control, and integration of data from different sources. Further, it lays the ground for development of supporting
tools and RegTech solutions.
Within the context of the EU AI Act, these supporting tools can be employed to assist \emph{AI providers} in documentation and exchange of information required for compliance, help \emph{conformity assessment bodies} in tasks related to auditing and certification, and assist \emph{AI deployers} in conducting fundamental rights impact assessment (FRIA). Further, AI Cards might become useful for the \emph{AI Office} in the development of automated 
Semantic Web-based tools to assist with FRIA, which has an overlap with technical documentation in terms of information requirements. Currently, there is little guidance for how to conduct risk assessment on the impact of AI systems on fundamental rights and the state of the art in taking fundamental rights seriously in the context of AI technical standardisation is new and fast evolving. Therefore, the locating, consumption and reuse of examples of AI risks assessment and accompanying technical documentation is urgently needed to establish a public knowledge base from which AI providers and deployers, especially less well-resourced SMEs and public bodies, can draw legal certainty. Such a sharing of AI risks and the documentation in AI card may therefore accelerate regulatory learning both between regulatory bodies and between prospective AI providers and deployers. Where confidentiality concern may impede the open access sharing of such information, official regulatory learning structures such as sandboxes and testing with human subjects could use this mechanism to maximise sharing and learning between participating stakeholders. Shared searchable repositories of AI risks and corresponding AI Cards may also offer support in identifying risks of reasonably foreseeable misuse of AI systems.

The semantic model, provided as an open-source ontology, can function as a basis for a pan-European AI vocabulary\footnote{See existing examples of EU vocabularies here: \url{https://op.europa.eu/en/web/eu-vocabularies}} to help establish a common language across different and multidisciplinary actors involved in the AI ecosystem. This consistency in the language, accompanied by a machine-readable representation, promotes interoperability and streamlines the information exchange required for incident reporting, compliance checking, and sharing best practices. It also promotes broader participation of stakeholders needed in development of standards for the AI Act \cite{veale2021demystifying}. In addition, it can be utilised for structuring the EU database of high-risk AI systems (Art. 60) and incident reports (Art. 62). As an added advantage, the semantic model can further be improved to evolve into a multilingual ontology supporting official EU languages. 

In a broader context, the semantic model is helpful for AI providers and users that operate in different jurisdictions in addressing challenges of cross-border compliance and interoperability by providing an extensible and adaptable structure to maintain information. Provided as an open resource, the schema can be reused and enhanced by the AI community to fit their needs in regard to documentation and risk management. 

\section{Conclusion and Further Work}

With this paper, we took an initial step to address the current lack of standardised and interoperable AI documentation practices in alignment with the EU AI Act. We proposed AI Cards, a novel framework for documentation of uses of AI systems in two complementary human- and machine-readable representations. We envision this contribution as a valuable input for future EU policymaking and standardisation efforts related to technical documentation, such as Implementing Acts, European Commission-issued guidelines, and European standards. We also hope that this work promotes use of standardised and interoperable Semantic Web-based formats in AI documentation.

In an ongoing effort to enhance AI Cards' usability and wider applicability, an in-depth user experience study (second phase of the evaluation) involving AI practitioners, auditors, standardisation experts, and policymakers is underway. 
Our future work also takes the direction towards development of automated tools for generating AI Cards from users' input and existing AI documents provided in textual formats. 
Further, alignment of AI Cards with documentation and reporting requirements of EU digital regulations, including the GDPR, Digital Services Act (DSA) \cite{dsa}, Interoperability Act \cite{interoperabilityact}, and Data Governance Act (DGA) \cite{dga}, as well as well-known risk management frameworks, e.g. NIST Artificial Intelligence Risk Management Framework (AI RMF 1.0) \cite{airmf}, is a key future step in refining the AI Cards.

\begin{credits}
\subsubsection{\ackname} 

This project has received funding from the European Union’s Horizon 2020 research and innovation programme under the Marie Skłodowska-Curie grant agreement No 813497 (PROTECT ITN), as part of the ADAPT SFI Centre for Digital Media Technology is funded by Science Foundation Ireland through the SFI Research Centres Programme and is co-funded under the European Regional Development Fund (ERDF) through Grant\#13/RC/2106\_P2.

\subsubsection{\discintname}
The views expressed in this article are purely those of the authors and should not, under any circumstances, be regarded as an official position of the European Commission.
\end{credits}
%

\bibliographystyle{splncs04}

\end{document}